\documentclass[a4paper,11pt]{article}
\pdfoutput=1 % if your are submitting a pdflatex (i.e. if you have
\usepackage{jinstpub} % for details on the use of the package, please
\usepackage{amsmath} 
\usepackage{xfrac}
\usepackage[capitalize]{cleveref}
\usepackage{xcolor} 
\usepackage{placeins}
\usepackage{latexsym}  

\renewcommand{\d}{\text{d}}

\newcommand{\Rinf}{R_\infty}
\newcommand{\Ia}{I_{\rm circ}}

\newcommand{\beq}{\begin{equation}}
\newcommand{\eeq}{\end{equation}}
\newcommand{\bq}{\begin{quotation}}
\newcommand{\eq}{\end{quotation}}

\begin{document}
\title{Intensity Limit in Compact H$^-$ and H$_2^+$ Cyclotrons}
\author{Thomas Planche, Richard A.\ Baartman, Hui Wen Koay, Yi-Nong Rao, Lige Zhang}
\affiliation{TRIUMF, 4004 Wesbrook Mall, Vancouver, B.C. Canada V6T 2A3}
\emailAdd{tplanche@triumf.ca}
\abstract{Compact H$^-$ cyclotrons are used all across the globe to produce medical isotopes. Machines with external ion sources have demonstrated average extracted currents on the order of a few mA, although reported operational numbers are typically around 1\,mA or below. To explore the possibility of extracting even more current from such cyclotrons, it is important to understand the mechanisms that drive intensity limits and how they scale. In this paper we review some of the key aspects of the beam dynamics in the central region of compact cyclotrons, including rf electric focusing and space charge effects. We derive the scaling of the phase acceptance with the rf gap voltage, harmonic number, etc. We also explore the scaling with different types of ions such as H$^-$, H$_2^+$ and H$_3^+$.  We discuss the impact of mechanical erosion of the central region electrodes. Thoughout the paper, we use examples and experimental data from two compact H$^-$ cyclotrons for reference: the TR-30 series and the TRIUMF 500\,MeV machine. }
\maketitle

\section{Introduction}
A cyclotron with a low injection energy, low enough that the radius of the first few turns is comparable to the magnet gap, is what we call a ``compact'' cyclotron. Such cyclotrons require no complicated pre-accelerator: they either use an internal ion source, or an external ion source with axial injection through a spiral inflector. But this advantage comes at a cost: in the central region the azimuthal field dependence, which is normally the source of vertical focusing, is smoothed out. 
%In that case, and for that region of the cyclotron, it is similar to the ``classic'' cyclotron originally invented by Ernest Lawrence. The salient feature is that there is essentially no magnetic vertical focusing. 
The only source of vertical focusing during the first few turns is the rf field. 
%As a result, the amount of beam that can be injected into a compact cyclotron is ultimately limited by a balance between vertical focusing from the rf gaps and the defocusing from space charge.
By contrast, in a ``ring'' cyclotron the beam is injected at a radius much larger than the magnet gap, and the azimuthal variation of the magnetic field insures vertical stability right from the first turn. In this respect, TRIUMF 500\,MeV cyclotron, despite its large size, has a very large gap ($\sim0.5$\,m) to accommodate the rf resonator and can perhaps paradoxically be regarded as a ``compact'' cyclotron.

H$^-$ cyclotrons present the significant advantage that they don't require individual turns to remain separated for extraction. The bunches can occupy the entire phase acceptance, and the intensity limit of these machines is essentially determined by the width of this phase acceptance. Compact H$^-$ cyclotrons are at the intersection: their phase acceptance is entirely dependent on the electric focusing from the rf gap in the central region.    

To study the intensity limit of these machines we start with a review of the rf-induced transverse focusing. Different formulas have been derived by several authors over the decades. We attempt to consolidate them with a formalism that combines the effects of arbitrary gap aspect ratio, large velocity change across one gap, and finite transit time, which are all relevant to existing high-current H$^-$ cyclotrons.

We continue with a detailed study of the phase acceptance in these machines in the absence of space charge, in which, along with electric focusing, geometrical considerations play a crucial role. We then quantify the reduction of phase acceptance caused by space charge, and arrive at the formula for the ultimate current limit in compact cyclotrons. In view of the recent regain of interest in compact cyclotrons to accelerate other particles than H$^-$, such as H$_2^+$ or H$_3^+$ molecules~\cite{meot2012mw,labrecque2013cyclotron,axani2016high,rao2019conceptual,rao2022innovative}, 
we pay a particular attention to the scaling of this limit with the mass and charge of the ion.
We cover longitudinal space charge and how it limits the bunching efficiency. We also cover what is a possibly less-known effect: the mechanical erosion of the central regions, which is found to be one of the most important limitations in TR-30 cyclotrons.

For context, let's recall that the mechanisms that limits the intensity in separated turn cyclotrons are fundamentally different. They are not covered in this paper. For a review of this particular topic read Ref.~\cite{baartman2013space} and the more recent contribution on the related beam break-up~\cite{cerfon2016vortex}.

\section{Electric Focusing}
\label{sec:electric-focusing}
%In cyclotrons, the transverse focusing from an rf gap is often referred to as `electric focusing'~\cite{wilson1938magnetic,reiser1971first,gordon1981electric}.  
In a compact cyclotron, various rf gap geometries may be encountered: the first, or the first few gaps may be square or oblong holes, while the rest display the more conventional `flat' dee-gap geometry.
The electric focusing depends on the aspect ratio: for reasons of symmetry, a gap as tall as it is wide will focus the beam as much in the horizontal direction as in the vertical direction, while an horizontally `flat' rf gap will focus only in the vertical direction.

Maxwell's equations impose that the amount of horizontal and vertical focusing follows a sum rule, as we are about to see. In this section we generalize the formula available in the literature for flat~\cite{wilson1938magnetic,reiser1971first} and cylindrical~\cite[section 7]{wangler2008rf}~\cite{TRI-DN-85-37} gaps, while paying particular attention to transit-time effects. For simplicity we exclude from our study the effect of asymmetric or tilted rf gaps which are discussed in Ref.~\cite{gordon1981electric,dutto1975focusing}. 

\subsection{Hamiltonian for Longitudinal and Transverse Motion}
Let's start from the general Courant and Snyder Hamiltonian~\cite[Appendix B]{courant1958theory} for a particle of mass $m$ and charge $q$, choosing for simplicity a straight reference trajectory:
\begin{equation}\label{eq:Hs}
    \mathcal{H}_s = -q A_s -\sqrt{\left(\frac{E-q\Phi}{c}\right)^2-m^2c^2-(P_x-qA_x)^2-(P_z-qA_z)^2}
\end{equation}
The canonical pairs of variables are $(x,P_x)$, $(z,P_z)$, $(t,-E)$, and $(s, \mathcal{H}_s)$, where $x$ and $z$ are the horizontal and vertical coordinates, and $E$ is the particle's total energy. The independent variable of this Hamiltonian is the longitudinal coordinate $s$.  %Additionally, $\Phi$ is the electric scalar potential and ${\bf A}$ is the magnetic vector potential.

First, one must choose a suitable set of potentials $\Phi$ and $\vec{A}$: along the optical axis the resulting electric field should be an arbitrary function $\mathcal{E}(s)$, there should be no magnetic field on axis, and Maxwell's equations should be satisfied in their homogenous form (no charge and no current on axis).  By analogy with the potential used in Ref.~\cite{Baartman:2017faa}, we propose Taylor series truncated to second order in $x$ and $z$:
\begin{equation}\label{eq:pots}
    \begin{aligned} 
        A_x  & = A_z = 0\,,                                                                                                                               \\ 
        A_s  & = -\mathcal{E}(s)\left(1 - \frac{\omega^2}{c^2} \frac{x^2 (1 + \iota) + z^2 (1 - \iota)}{4}\right) \frac{\sin(\omega t + \phi)}{\omega}\,, \\ 
        \Phi & = \mathcal{E}'(s) \cos(\omega t + \phi) \frac{x^2 (1 + \iota) + z^2 (1 - \iota)}{4}\,. 
    \end{aligned}
\end{equation}
The ``aspect-ratio'' parameter $\iota$ can take any value between -1 and +1; $\iota = 0$ corresponds to the case of a round gap~\cite{Baartman:2017faa} while $\iota = -1$ corresponds to the case of a horizontal gap. Intermediate values of $\iota$ can be used to treat intermediate geometries such as oblong gaps. 

With $\iota$ assumed to be constant, one can verify that this set of potentials leads to the desired fields:
\begin{align}
    \vec{\mathcal{E}}(0,0,s) & = -\nabla\Phi - \frac{\partial \*A}{\partial t} = (0,0,\mathcal{E}(s)) \,, \\
    \vec{B}(0,0,s)           & = \nabla \times \*A = (0,0,0)\,,
\end{align}
while at the same time satisfying Maxwell's four equations in their homogenous form along the reference trajectory ($x=z=0$). 
% \begin{align}
%     \nabla \cdot \vec{B}                                                                & = 0\,,        \\
%     \nabla \cdot \vec{\mathcal{E}}                                                      & = 0\,,        \\
%     \nabla \times \vec{B} - \frac{1}{c^2} \frac{\partial \vec{\mathcal{E}}}{\partial t} & = (0,0,0)\,,  \\
%     \nabla \times \vec{\mathcal{E}} + \frac{\partial \vec{B}}{\partial t}               & =  (0,0,0)\,.
% \end{align}
Remember that the choice of potentials is not unique; suitable choices differ by a gauge transformation. %With this particular choice of gauge, the electric potential $\Phi$ is zero on axis; The on-axis energy gain comes entirely from the time derivative of $A_s$.

\subsection{Longitudinal Motion}
From~\cref{eq:Hs} we  obtain the longitudinal equations of motion for the reference particle ($x=z=0$ and $P_x=P_z=0$):
\begin{align}\label{eq:eomt}
    t' & = -\frac{\partial \mathcal{H}}{\partial E} = \frac{E}{c \sqrt{E^2-m^2c^4}} = \frac{1}{\beta c} \,, \\[0.5em] \label{eq:eome}
    E' & = +\frac{\partial \mathcal{H}}{\partial t} = q \mathcal{E} \cos(\omega t +\phi) \,,
\end{align}
where the symbol $'$ denotes a total derivative with respect to~$s$.
By choosing $t=0$ at the center of the gap, one can write the total energy gain through the gap as:
\begin{equation}
    \Delta E = q V_\text{g} \mathcal{T} \cos\phi\,,
\end{equation}
where $V_\text{g}=\int_{-\infty}^{\infty}\mathcal{E}(s)\d s\ne0$ is the gap voltage, and the transit-time factor $\mathcal{T}$ is given by:
\begin{equation}\label{eq:tof}
    \mathcal{T} = \frac{1}{V_\text{g}}\int_{-\infty}^{\infty}\mathcal{E}(s)\cos(\omega t(s))\d s\,.
\end{equation}
See for instance Ref.~\cite[section 2.2]{wangler2008rf}.

%The signs come from the fact that the time coordinate $t$ is canonically conjugate with $-E$.
% Note that the definition of phase that we use here leads to an energy gain per unit length proportional to $\cos\phi$. Some authors prefer to use the other $\sin\phi$ convention, but we adopt here the cosine convention that seam more common in the cyclotron literature.
% Some authors prefer to use the other convention, where it is proportional to the \textbf{sine} of the phase: beware!

\subsection{Linear Transverse Motion and Approximate Thin-Lens Formula}
To study the transverse motion close to the optical axis, we Taylor expand the Hamiltonian to second order in $x$, $z$, $P_x$, and $P_z$, leading to:
\begin{equation}\label{eq:H2}
    \mathcal{H}_2(x,P_x,z,P_z;s) = \frac{P_x^2}{2P}+\frac{P_z^2}{2P}+\frac{q}{2}\left(\frac{\mathcal{E}'C}{\beta c}-\frac{\mathcal{E}S \omega}{c^2}\right)\frac{x^2 (1 + \iota) + z^2 (1 - \iota)}{2}\,,
\end{equation}
where $C = \cos(\omega t + \phi)$, $S = \sin(\omega t + \phi)$, and $P = \sqrt{\frac{E^2}{c^2}-m^2c^2}$ is the reference particle's momentum.
From this quadratic Hamiltonian we obtain the equation for linear transverse motion:
\begin{equation}\label{eq:lintran}
    P_z' = -\frac{\partial \mathcal{H}_2 }{\partial z} = z\,q\frac{1 - \iota}{2 c} \left(\mathcal{E}S\frac{\omega}{c} - \frac{\mathcal{E}'C}{\beta } \right)\,.
\end{equation}
Note that the vertical coordinate $z$ can be swapped for the horizontal $x$ by changing the sign in front of the aspect-ratio parameter~$\iota$.
% This expanded Hamiltonian contains no first order term: it implies that a particle placed on the reference axis will remain on it. The $0^\text{th}$ order term was dropped as it does not contribute to the dynamics.
For any given on-axis electric field $\mathcal{E}(s)$\footnote{Not any: $\mathcal{E}(s)$ must be at least differentiable.}, the linear transport can be calculated numerically, to arbitrary precision, by integrating~\cref{eq:lintran} along with~\cref{eq:eomt,eq:eome}. This approach is implemented for instance in the envelope code \texttt{TRANSOPTR}\,\cite{Baartman:2017faa}.

Let us however trade a little precision for some more intuitive comprehension by deriving the approximate integral using the following simplifying assumptions.
\begin{enumerate}
  \item The thin-length approximation: the gap is sufficiently short that we can ignore the variation of $z$ across it. This is usually the case in the central region of compact cyclotrons, or close enough\,\cite{reiser1971first}.   
  \item The hard-egde approximation: let $\mathcal{E}(s)$ be a top-hat function of
  effective length as $L_{\rm eff}:=\frac{V_\text{g}}{\mathcal{E}(0)}$.
\end{enumerate}
Using the first assumption, we integrate~\cref{eq:lintran} for an incident parallel ray:
\begin{equation}\label{eq:384}
    \Delta P_z  = z\,q\frac{1 - \iota}{2 c}\int_{-\infty}^\infty\left(\mathcal{E}S\frac{\omega}{c} - \frac{\mathcal{E}'C}{\beta } \right)\,\text{d}s\,.
\end{equation}
Since the top-hat function we have chosen for $\mathcal{E}$ is discontinuous, we must transform out $\mathcal{E}'$ before performing the integration. This is done by noting that:
\begin{equation}
    \frac{{\cal E}'C}{\beta}\d s=\d\left(\frac{{\cal E}C}{\beta}\right)-{\cal E}\left(\frac{C}{\beta}\right)'\d s\,.
\end{equation}
The first term in this expression integrates to zero as ${\cal E}=0$ at both ends of the integral. Further, $(C/\beta)'=C'/\beta-C\beta'/\beta^2$, $C'=-\omega S/(\beta c)$. The result is
\begin{equation}
    \label{eq:386}\int\frac{{\cal E}'C}{\beta}\d s=\frac{\omega}{c}\int\frac{{\cal E}S}{\beta^2}\d s+\int\frac{\beta'}{\beta^2}{\cal E}C\d s.
\end{equation}
We can now identify the first term in~\cref{eq:384} as $-\beta^2$ times the first term here in~\cref{eq:386}, and thus see that the former is the effect of the rf magnetic field. Ordinarily in this non-relativistic regime for cyclotron injection, we could ignore it, but for completenes, we leave it. The sum of the first term in~\cref{eq:384} and the first term in~\cref{eq:386} contributes a factor $\frac{1}{\beta^2}-1=\frac{1}{\beta^2\gamma^2}$. 
%, so
% \begin{equation}
%     \int\left(\mathcal{E}S\frac{\omega}{c} - \frac{\mathcal{E}'C}{\beta } \right)\,\text{d}s=-\frac{\omega}{c}\int\frac{{\cal E}S}{\beta^2\gamma^2}\d s-\int\frac{\beta'}{\beta^2}{\cal E}C\d s.
% \end{equation}
Using $\beta'=\frac{\gamma'}{\beta\gamma^3}=\frac{q{\cal E}C}{\beta\gamma^3 mc^2}$, our integral becomes:
\begin{equation}
    \int\left(\mathcal{E}S\frac{\omega}{c} - \frac{\mathcal{E}'C}{\beta } \right)\,\text{d}s=-\frac{\omega}{c}\int\frac{{\cal E}S}{\beta^2\gamma^2}\d s-\frac{q}{mc^2}\int\frac{({\cal E}C)^2}{\beta^3\gamma^3}\d s.
\end{equation}

These integrals can be approximated using the hard-edge approximation, and setting the value of $\beta\gamma$ to be constant and equal to the average between the value before and after crossing the gap, leading to:
%We also approximate $\mathcal{E}(s)$ with a top-hat function of
% effective length as $L_{\rm eff}:=\frac{V_\text{g}}{\mathcal{E}(0)}$.
% \begin{equation}
%     \int({\cal E}C)^2 \d s = (V_\text{g}\widetilde{T}\cos\phi)^2 + \mathcal{O}(\Delta \phi)^4\,,
% \end{equation}
% The integrated vertical momentum kick becomes:
\begin{equation}
    \label{eq:389}
    \boxed{\Delta P_z = -z\frac{1-\iota}{2c}\left[\frac{\omega}{c}\frac{qV_\text{g}\mathcal{T}\sin\phi}{\beta^2\gamma^2}+\frac{(qV_\text{g}\widetilde{\mathcal{T}}\cos\phi)^2}{L_{\rm eff}mc^2\beta^3\gamma^3}\right]\,.}
\end{equation}
The term proportional to $V_\text{g}\mathcal{T}\sin\phi$ is the pure rf part, including the effect from the electric and magnetic time-varying fields. The transit time factor $\mathcal{T}$ from~\cref{eq:tof} becomes in the hard-edge approximation:
\begin{equation}\label{eq:t-he}
    \mathcal{T} = \frac{1}{L_{\rm eff}}\int_{-L_\text{eff}/2}^{L_\text{eff}/2}\cos(\omega t(s))\d s= \frac{\sin{\delta \phi}}{\delta \phi}\,,
\end{equation}
where $\delta \phi = \frac{L_\text{eff} \omega}{2 \beta c}$ is half of the time taken by the particle to cross the gap measured in unit of rf phase.

The term proportional to $(V_\text{g}\cos\phi)^2$ represents what is called ``electrostatic focusing'' in Ref.~\cite[section 7]{wangler2008rf}. It is identical to that produced by a DC gap of the same voltage. 
% It comes from the integral of the change in velocity $\beta'$.
Unlike for the pure rf term, transit time effects on the electrostatic term is phase-dependent:
\begin{equation}\label{eq:ttilde}
    % \widetilde{\mathcal{T}}^2 = 1-2(1-\mathcal{T})(1-\tan^2\phi)\,.
    \widetilde{\mathcal{T}}^2 = 1 - \frac{1}{2}(1 - \mathcal{T} \cos \delta\phi)(1-\tan^2\phi)\,,
\end{equation}
If we expand~\cref{eq:t-he,eq:ttilde} for small values of $\delta\phi$ we find that:
\begin{align}
  \mathcal{T} &= 1-\frac{\delta\phi^2}{6} +\mathcal{O}(\delta\phi^4)\,,\\
  \widetilde{\mathcal{T}}^2 &= 1-\frac{\delta\phi^2}{3}(1-\tan^2\phi) +\mathcal{O}(\delta\phi^4)\,.
\end{align}
This means that near the crest ($\phi=0$), that is the main region of interest in upcoming~\cref{sec:phase_acceptance}, the electrostatic focusing diminishes more rapidly than the pure rf focusing for larger transit times. As we are about to see, the electrostatic term is critical as it extends the phase acceptance on the $\phi<0$ side of the crest and affects the intensity limit in compact cyclotrons. \cref{eq:389} shows that this term depends strongly on the effective length of the rf gap, not only through the $\frac{1}{L_\text{eff}}$ dependence, but also because of the increased sensitivity to transit time. To maximize the phase acceptance of a cyclotron, it is critical to make the effective length of the rf gap in the central region as short as possible. 
% This fact may however be counteracted by the ``strong-focusing'' effect between the entrance and exit edges of the gap, that we have neglected as a result of assumption 1. This is expected to be proportional to $L_\text{eff}$, and is most accurately quantified with a full numerical integration.

To simplify formulas we set from now on $\iota$ to -1 and $\mathcal{T}$ to 1. But one can always come back to~\cref{eq:389} to generalize equations hereunder to include the effect of gap aspect ratio and transit time.

\subsection{Vertical Tune at the Centre of a Compact Cyclotron}
For the cyclotron application, $\omega=\beta ch/R$ where $h$ is the harmonic number, i.e.~the ratio of rf frequency to revolution frequency. Further, at injection $\beta\ll 1$ and we set $\gamma=1$. Let there be $n_{\rm g}$ rf gaps per turn; we assume for simplicity that the bunch phase $\phi$ is the same at each gap, though this can be generalized. The $\Delta\theta=2\pi/n_{\rm g}$.
%Except for the first rf gap, which is often close to being round ($\iota=0$), the rest of the rf system is normally flat and focuses only in the $z$ direction ($\iota=-1$). 
The smoothed betatron motion equation $z''+\nu_z^2z=0$ gives $\nu_z^2z=-\frac{R}{P}\frac{\d P_z}{\d\theta}$. \cref{eq:389} can thus be transformed into an equation for vertical tune:

\begin{equation}
    \label{eq:nuz2EF}
    \nu_{z\text{e}}^2=-\frac{R}{zP}\frac{\Delta P_z}{\Delta\theta}=n_{\rm g}\frac{1}{4\pi}\left[h\left(\frac{qV_\text{g}\mathcal{T}\sin\phi}{E_{\rm k}}\right)+\frac{R}{2 L_{\rm eff}}\left(\frac{qV_\text{g}\widetilde{\mathcal{T}}\cos\phi}{E_{\rm k}}\right)^2\right]\,.
\end{equation}
$E_{\rm k}$ is the kinetic energy.
\begin{figure}[ht] 
    \begin{center}
        \includegraphics[width=0.7\textwidth]{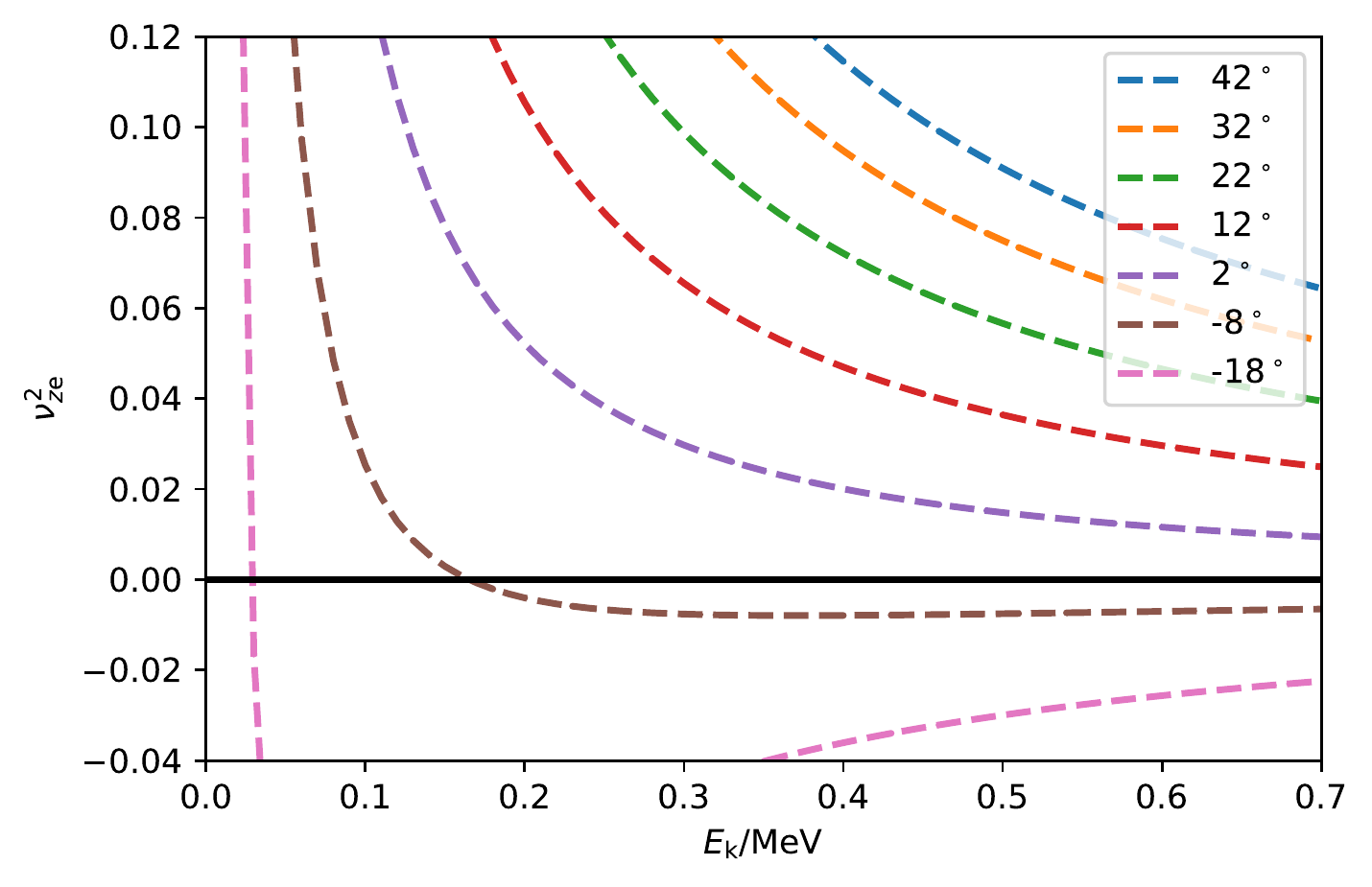}
        \caption{Electric focusing contribution in the TR-30 cyclotron to $\nu_z^2$ for seven phases (from lowest to highest curves) $-18^\circ$ to $+42^\circ$ in steps of $10^\circ$. The parameters for this cyclotron are~\cite{baartman1995intensity,rao2022innovative}: $\Rinf=$~2.6\,m, $h = 4$, $n_{\rm g} = 4$ gaps per turn, $V_\text{g} = 50$\,kV and $L_{\rm eff} = 1.3$\,cm. This is a typical example, where electric $\nu_z\sim0.2$ on the first few turns.}\label{fig:TR-30electricf}
    \end{center}
\end{figure}
As the vertical tune varies with phase, the resulting beam has a heavily nonlinear character. The transverse acceptance ellipse has an aspect ratio that varies along the length of the injected bunch, which result in transverse bunch profiles such as the ones shown in~\cite[Fig.~4.8]{root1974experimental} and~\cite[Fig.~3]{blackmore1973experimental}.

Typically, the voltage across the acceleration gaps is comparable with the energy of the injected beam. The surviving particles are the ones with the most acceleration and these have phase near zero. Thus on the first gap the second term dominates and thereafter drops quickly compared with the first term. An example for the compact cyclotron TR-30 is shown in Fig.\,\ref{fig:TR-30electricf}. If not for the second term, only positive phases would be focussed. These are the phases on the falling part of the acceleration. In this example, phases larger than $-5^\circ$ are focussed.
This, together with the minimum energy gain needed to avoid the injection gap after one turn, establishes the phase acceptance, as we are about to study in detail in the following section.

\section{Phase Acceptance}\label{sec:phase_acceptance}
\subsection{Without Space Charge}
% To make it into the ``phase acceptance'' of a compact cyclotron, a particle must on the one hand gain enough energy on the first turn to miss the injection gap.

For an on-crest particle to complete the first turn without intercepting the injection gap, the voltage must exceed a threshold  $V_\text{m}$. Below this threshold, no particle can be injected into the cyclotron, i.e.~the phase acceptance is zero. The exact value of $V_\text{m}$ depends on the geometry of the central region.
For larger values of the gap voltages $V_\text{g}$  the edge of the phase acceptance moves away on both sides of the crest as:
\begin{equation}\label{eq:pinchoff}
    \phi =  \pm\arccos\left(\frac{V_\text{m}}{V_\text{g}}\right)\,.
\end{equation}

A particle must also receive enough vertical focusing from the rf gaps to survive until it reaches larger radii where the azimuthally varying focusing kicks in.  Since the electric focusing is phase dependent, any phase for which $\nu_z^2$ remains negative ``long enough'' will fall outside of the acceptance. The exact position of the edge of the vertical acceptance depends on  the vertical beam emittance and aperture size (see~\cite[Eq.~5]{baartman1995intensity}), but for simplicity we will study the limit case of a zero-vertical-emittance beam,  and places the edge of the acceptance exactly at $\nu_z^2=0$. The corresponding phase is generally close enough to the crest that we can expand~\cref{eq:nuz2EF} in Taylor series around $\phi=0$, and to first order we find that the edge of the phase acceptance is given by:
% \begin{equation}\label{eq:nuzphi}
%     \phi = \frac{q(1-2T)}{2 h}\frac{R}{L_\text{eff}}\frac{V_\text{g}}{E_\text{k}}+\mathcal{O}(\phi^3)\,.
% \end{equation}
\begin{equation}\label{eq:nuzphi}
    \begin{aligned}
        \phi_\text{edge} &\approx -\kappa V_\text{g}\,,\\
        \kappa &= \frac{q\Rinf}{h mc^2\beta L_\text{eff}}\,,
    \end{aligned}
\end{equation}
%$\kappa = \frac{2\mathcal{T}-1}{\mathcal{T}} \frac{qR}{2 h E_\text{k} L_\text{eff}}$
where the values of the parameters $\beta$ and $L_\text{eff}$ are taken at the radius of the vertical acceptance ``bottleneck'', i.e.~the radius where $\nu_{z{\rm e}}^2+\nu_{z{\rm m}}^2$ goes through a minimum, see~\cref{fig:phaseplot-TRIUMF}.
%between radially decreasing electric focusing (\cref{fig:TR-30electricf}) and emerging AVF focusing. 
Combining~\cref{eq:pinchoff} and~\cref{eq:nuzphi} we find that the phase acceptance, without considering space charge, is given by:

\begin{equation}\label{eq:Deltaphi}
    \Delta \phi = \arccos\left(\frac{V_\text{m}}{V_\text{g}}\right) + \min\left(\arccos\left(\frac{V_\text{m}}{V_\text{g}}\right), \,\,\kappa V_\text{g}\right)\,.
\end{equation}

\begin{figure} 
    \begin{center}
        \includegraphics[width=0.75\textwidth]{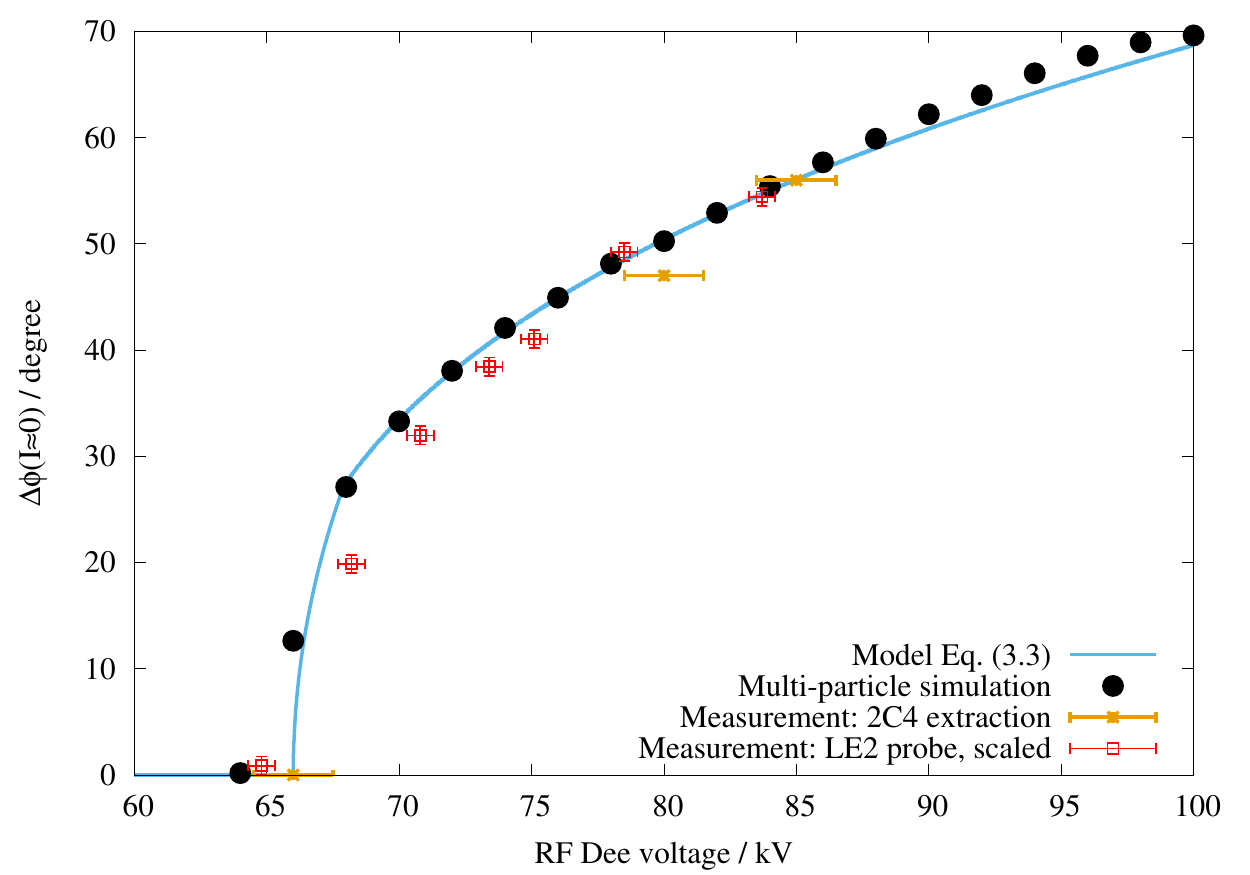}
        \caption{Phase acceptance, in the limit of low space charge, as a function of the rf dee voltage for the TRIUMF 500\,MeV cyclotron. Prediction from~\cref{eq:Deltaphi} is compared with actual measurements using an unbunched beam, as well as multiparticle simulations without space charge. Two different measurements were taken on two different days, using different diagnostics equipment to measure beam current and rf dee voltage, hence the differences in error bars. The phase acceptance was measured as the ratio between in injected cw beam current, and the current measured either on a radial probe (LE2) or down an extraction beamline (2C4). Note that the measurement taken using the LE2 is scaled up by 20\% to account for the fact that the probe is not designed as a proper Faraday cup, and so does not collect all the charge.}\label{fig:phaseplot}
    \end{center}
\end{figure}

This simple model was tested, both experimentally and using simulations, on the TRIUMF 500\,MeV cyclotron. The threshold dee voltage below which no beam gets past the first turn is found experimentally to be 66\,kV. 
We also measured the phase acceptance at low space charge by injecting an unbunched beam for several values of the dee voltage. The data is presented in~\cref{fig:phaseplot}, along with results from  multiparticle simulation, without space charge, using the code \texttt{TRIWHEEL}~\cite{TRI-BN-16-21}. The theoretical prediction from~\cref{eq:Deltaphi} fits best both experimental and simulation results with a value of $\kappa \approx 20^\circ/100\,\text{kV}$.

% We also measured the value of the phase acceptance at low space charge for a particular value of the rf dee voltage, by injecting an unbunched beam, and found that $\kappa \approx 20^\circ/100\,\text{kV}$.  We repeated this measurement for different values of the dee voltage and compared the result with the prediction from~\cref{eq:Deltaphi}, see~\cref{fig:phaseplot}. We also compare the same results with prediction from multiparticle simulation, without space charge, using the code \texttt{TRIWHEEL}~\cite{TRI-BN-16-21}.

Since $\nu_z$ goes through a minimum around 1\,MeV, see~\cref{fig:phaseplot-TRIUMF}, evaluating~\cref{eq:nuzphi} with $h=5$ and $\Rinf = 10$\,m leads to $\kappa$ of $20^\circ/100\,\text{kV}$ for an effective length of the rf gap of $\sim$1\,inch. The physical gap length is actually 2\,inch at the corresponding radius~\cite{TRI-BN-16-21}. The discrepancy may be explained by the effect of gap geometry that we have neglected so far, and which had been carefully optimized by the original designers of the cyclotron~\cite{dutto1975focusing}. 
We have in particular neglected the variation of the form factor $\iota$ across the gap: if $\iota$ is a function of $s$, the potentials in~\cref{eq:pots} would have to take a different form to satisfy Maxwell's equations. Several authors have taken advantage of a varying form factor to increase the vertical focusing on the first few gaps, see for instance~\cite[section 3.6]{root1974experimental} and~\cite[section 6]{rao2022innovative}.
  
Both the experimental and multiparticle simulation results show a somewhat smoother drop of the acceptance around the threshold voltage than predicted by our model: this is due to the fact that we have neglected the effect of the non-zero radial beam emittance. Otherwise, the predicted power of this simple model is good, and is used routinely to assess the machine performance during operation, as the rf dee voltage is often not constant throughout the running year, and for technical reasons is run anywhere between 82 and 95\,kV.

But the validity of this model diminishes with increasing amount of current injected into the machine: this is due to the effect of space charge that we are about to discuss.

\subsection{With Space Charge}
Space charge produces a vertical defocusing force that shifts the value phase for which $\nu_z$ drops to zero, reducing the phase acceptance accordingly. Working from the Laslett's formula~\cite{laslett1963intensity}, the incoherent tune shift in the central region of a cyclotron can be approximated by~\cite{{baartman1995intensity}}:
\begin{equation}\label{eq:dnu2sc}
    \Delta(\nu_z^2)_\text{sc}\approx\frac{-4}{\beta}\,\frac{\hat{I}}{I_0}\,\left(\frac{\Rinf}{b}\right)^2,
\end{equation} where $b$ is the vertical (2 rms~\cite{sacherer1971rms}) beam size, and $I_0 = \frac{m c^2}{q 30\Omega}$ is a normalized current introduced by Joho~\cite{joho1981high}, where $30\,\Omega$ is the impedance of free space divided by $4\pi$. Note that this formula is for the non-relativistic regime, otherwise one must also divide by $\gamma^3$.

Considering again the limit case of a zero-vertical-emittance beam, the edge of the phase acceptance is found where $\nu_z^2 = \nu_{z\text{e}}^2+\Delta(\nu_z^2)_\text{sc} = 0$, which leads to the version of~\cref{eq:nuzphi} including space charge:
\begin{equation}\label{eq:nuzphisc}
    \begin{aligned}
        \phi_\text{edge} & \approx \xi\hat{I} -\kappa V_\text{g} \,,\\
        \xi  & = \frac{2R^2Z_0}{b^2\beta h n_\text{g}V_\text{g}}\,,
    \end{aligned}
\end{equation}
% \begin{equation}\label{eq:nuzphisc2}
%     \phi_\text{edge} = \frac{R}{h\beta}\left(\frac{R\, \hat{I}\, Z_0}{b^2 n_\text{g}V_\text{g}} -\frac{1}{m \beta }\frac{q V_\text{g}}{c^2 L_\text{eff}} \right)\,.
% \end{equation}
where $Z_0 \approx 4\pi \times 30 \Omega$ is the impedance of free space. This new parameter $\xi$ is of utmost importance: it quantifies the loss of phase acceptance per mA of \textbf{peak} beam current. The fact that space charge and electric focusing oppose each other is apparent in the fact that $\xi\hat{I}$ and $-\kappa V_\text{g}$ have opposite signs. The former shifts the edge of the phase acceptance towards positive values while the later shifts it towards negative values of $\phi$.

We can finally re-write~\cref{eq:Deltaphi} to include the effect of space charge:
\begin{equation}\label{eq:Deltaphi2}
    \Delta \phi(\hat{I}) = \arccos\left(\frac{V_\text{m}}{V_\text{g}}\right) + \min\left(\arccos\left(\frac{V_\text{m}}{V_\text{g}}\right), \,\,\kappa V_\text{g} - \xi\hat{I}\right)\,.
\end{equation}

\begin{figure}
    \begin{center}
        \includegraphics[width=0.75\textwidth]{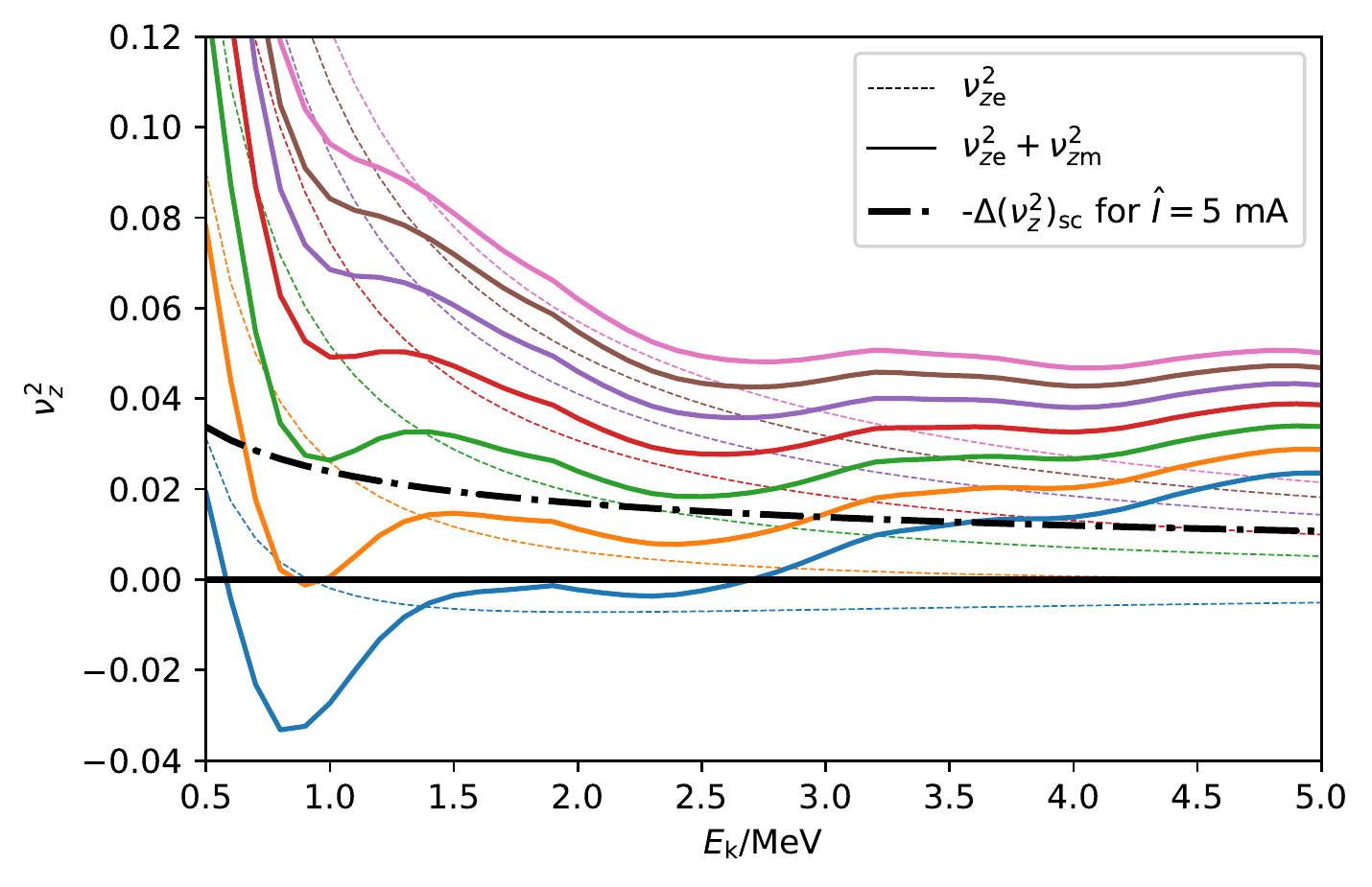}
        \caption{Vertical tune square plotted for rf phases ranging from 
        -18$^\circ$ to +42$^\circ$ in streps of 10 degrees. The bottom (blue) lines correspond to the -18$^\circ$ case, while the top (pink) curves correspond to the +42$^\circ$ case. $\nu_{z\rm e}^2$ is calculated using~\cref{eq:nuz2EF} with a dee voltage $=V_\text{g}/2=$~90\,kV, $n_\text{g}=$2, $h=5$ and $L_\text{eff}=$~1\,inch. This is approximate since in fact $L_\text{eff}$ changes with radius. $\nu_{z\rm m}^2$ is calculated using the orbit code \texttt{CYCLOPS}. Note that $\nu_{z\rm m}^2<0$ below $\sim$1.3\,MeV. This plot is comparable with~\cite[Fig.\ 2]{dutto1972optimization}.  
        }\label{fig:phaseplot-TRIUMF}
    \end{center}
\end{figure}

For TRIUMF cyclotron, we can estimate the 2rms vertical beam size in the central region to be $b \approx 0.3$ inch, using measured data such as what is presented in~\cite[Fig.~10]{planche2013measurement}. Evaluating again for an energy of 1\,MeV and for a dee voltage of 90\,kV gives $\xi\approx 1.9^\circ$/mA. This value can be compared with the experimentally measured value of $\approx 2^\circ$/mA reported in Ref.~\cite{dutto1988upgrading}. We also reproduced the calculation shown in~\cite[Fig.~2]{dutto1972optimization} using~\cref{eq:nuz2EF} to calculate the electric focusing, and a \texttt{CYCLOPS} run with the historical magnetic survey data CYC581~\cite{zhang2022asurement}. The result is presented in~\cref{fig:phaseplot-TRIUMF}. 

Superimposed on this figure is a line that shows the vertical tune shift caused by a 5\,mA peak current: this shows that about 10$^\circ$ of phase acceptance is expected to be lost at this current, which supports a value of $\xi$ around 2 degree/mA.  

% One may argue that, at the edge of the phase acceptance, $\Delta(\nu_z^2)_\text{sc}$ is actually only about \textonehalf~of what is given by~\cref{eq:dnu2sc}, since the space charge forces at the edge of the beam are only  \textonehalf~of that for a 2-dimentional DC beam. But the point of this discussion is not so much to derive accurate quantitative predictions: this is better done with detailed 3-dimensional simulations. What we are looking for here are predictions in terms of scaling of the effects of space charge with respect to parameters such as the rf gap voltage or the particle mass and charge. 

\FloatBarrier
\subsection{Absolute Average Current Limit}
The average current is the product of the peak current $\hat{I}$ times the acceptance:
\begin{equation}\label{eq:Iav}
    I = \hat{I}\frac{\Delta \phi}{2\pi}\,.
\end{equation}
The limit $\hat{I}\rightarrow 0$ leads to the maximum phase acceptance given by~\cref{eq:Deltaphi}, but leads to zero average beam current. On the other extreme $\hat{I}\rightarrow \infty$ leads to zero phase acceptance, and so zero average beam current as well. Somewhere in between is a value of $\hat{I}$ which leads to the maximum possible current output: this is the current limit we have been looking for.

Without too much loss of generality\footnote{Excluding only the cases with $V_\text{g}\approx V_\text{m}$ where consideration of vertical focusing are irrelevant.}
we can simplify~\cref{eq:Deltaphi2} into:
\begin{equation}\label{eq:Deltaphi3}
    \Delta \phi(\hat{I}) = \arccos\left(\frac{V_\text{m}}{V_\text{g}}\right)  \,\,+ \kappa V_\text{g} - \xi\hat{I} = \Delta \phi(0) - \xi\hat{I}\,.
\end{equation}
Combined with~\cref{eq:Iav} we find that the maximum average current is achieved for $\hat{I}$ such that half of the zero-current phase acceptance is lost. The maximum achievable average beam current is given by:
\begin{equation}
    I_\text{max} = \frac{1}{2\pi\xi} \left(\frac{\Delta \phi(0)}{2}\right)^2\,,
\end{equation}
or more explicitly:
\begin{equation}\label{eq:curlim}
    I_\text{max} = \frac{b^2\beta h n_\text{g}V_\text{g}}{4\pi R^2Z_0} \left(\frac{\Delta \phi(0)}{2}\right)^2\,.
\end{equation}
% This formula allows to determine that the empirical parameter or $\left(\frac{1}{56^\circ}\right)$\,mA reported on~\cite[Sec.~3]{dutto1988upgrading} corresponds to a value of $\xi\approx 2.2^\circ/\text{mA}$, once again consistent with the range of 2 to 3$^\circ/\text{mA}$ discussed at the end of the previous section. 

This formula allows to predict an ultimate current limit for the TRIUMF cyclotron, using the conservative value of $\xi$ of 3$^\circ$/mA, of \textbf{1.1\,mA} for a dee voltage of 100\,kV. The highest current output ever reported is 0.4\,mA~\cite[Fig.~1]{dutto1988upgrading} (measured at a reduced duty cycle). This discrepancy is explained by the fact that the brightness of the ion source and the bunching system together have never allowed to reach over $\hat{I}\approx 2.5$\,mA, far below the 12\,mA required to reach this limit. 

Using the same formula with the parameters of the TR-30 cyclotrons~\cite[Tab.\ 1]{rao2022innovative}, assuming that the minimum of vertical focusing happens around 1\,MeV (see~\cref{fig:phaseplot}), using $\Delta\phi(0)=60^\circ$ from experimental data~\cite[Fig.\ 1]{baartman1995intensity}, and using a vertical beam size $b =$~5\,mm as in~\cite[section 3]{baartman1995intensity}, we find an absolute current limit of \textbf{3.7\,mA}. This is consistent with the empirical current limit of 3.3\,mA reported in Ref.~\cite{baartman1995intensity}. 

Remember: this is the current limit for compact cyclotrons that use stripping for extraction. The mechanisms that limit the intensity in separated-turn cyclotrons are entirely different~\cite{baartman2013space}. In particular the famous $I_\text{max}\propto V_\text{g}^3$ discussed in~\cite[section 7]{joho1981high} does not apply to stripping extraction.

\subsection{Space Charge Scaling for H$_2^+$ vs.\ H$^-$ Ions}
Let's consider two hypothetical cyclotrons, one accelerating H$_2^+$ ions, the other H$^-$. Let these two machines have the same central magnetic field $B_\text{c}$, a similar magnetic field structure (same gap height, and sector geometry) and the same number of rf gaps and gap voltage.  The same $B_\text{c}$ means that $\Rinf = \frac{mc}{q B_c}$ is twice as large for H$_2^+$ and for H$^-$. Since in a cyclotron $R = \beta \Rinf$, this means that the velocity of H$_2^+$ is half of that of H$^-$ at a given radius. A similar magnetic structure means that the magnetic focusing $\nu_{z\rm m}$ dependence on radius is the same, since the geometry of the closed orbits will be identical~\cite{planche2019designing}. This suggests that comparison between the two machines should be done at a given radius, and not at a given velocity as in Ref.~\cite{baartman2015space}
%This means in particular the the AVF focusing will start to ``kick in'' at the same radius, and that the vertical acceptance bottleneck will be at about the same radius for both machines.

% Comparison for a fixed value of $\beta$ , which is consistent with Ref.~\cite{baartman2015tri} 

At a given radius then, the lower velocity makes the space charge effect stronger for H$_2^+$ compared to H$^-$: since $\xi\propto \frac{1}{\beta}$, the phase acceptance shrinks twice as fast with increasing peak currents. But for the same reason, the strength of the electric focusing is also stronger, with $\kappa$ twice as large. The two effects cancel each other exactly for a gap voltage:
\begin{equation}
    V_\text{g} = \sqrt{\frac{2\hat{I}L_\text{eff}\Rinf Z_0\beta^2 mc^2}{qb^2n_\text{g}}}\,.
\end{equation}
For larger gap voltages the phase acceptance is larger for the H$_2^+$ machine; for lower voltages, it is larger for the H$^-$ machine.

Additionally, for a given circulating electrical peak current $\hat{I}$, the extracted proton current is twice as much from an H$_2^+$ machine. Using a sufficiently large rf gap voltage, the acceleration of H$_2^+$, or even H$_3^+$, in a compact cyclotron has from this perspective the potential to produce higher proton currents than the with H$^-$ ions.

\section{Bunching}
H$^-$ ion sources of $>20$\,mA and small emittance exist so in principle compact cyclotrons should easily be able to exceed 2\,mA output beam. A DC beam is desirable, since it allows near 100\% space charge neutralization, and no intensity-dependent space charge detuning in the injection line. But this requires most of the $\sim20$\,mA to be dumped onto the first turn. As shown in the next section, this approach too quickly erodes the central region components. 

To minimize erosion, it would be beneficial to bunch the beam into the cyclotron acceptance window. But here the small size and low injection energy of the compact cyclotron make this hardly practicable. Typically the rf frequency is 4 times the revolution frequency, which for a 1.2\,T central field is 18\,MHz, while the injection energy is 25\,keV. Thus, the spatial rf frequency in the injection line is $\beta\lambda_{\rm rf}=3$\,cm, and a double-gap buncher has a distance of 15\,mm between gaps; a distance that would be similar to its aperture. This results in transit time factor that varies across the aperture by over a factor of 2. 

Further, longitudinal space charge works against the buncher. There is a characteristic length $L_0$ that determines the rate of expansion with drift distance of a bunch\cite{baartman1995intensity}:\begin{equation}\label{l0}
L_0\equiv\sqrt{\frac{V_0}{30\,\Omega\times\Ia}\,\frac{r_0^3}{\lambda_{\rm rf}}},
\end{equation}
where $V_0$ is particle energy per charge, i.e.\ the net acceleration voltage from the ion source. Note that $L_0$ is independent of the particle mass and so applies to any ion as long as $\Ia$ is the electric current, not particle current. Starting at the first accelerating gap in the cyclotron and working backwards to find the bunching conditions, this assumes a spherical bunch of radius $r_0$. The spherical shape is an approximation, but the scaling as inverse square root of the space charge density applies generally. Multiplied by an expansion factor $r/r_0=2\pi/\Delta\phi$ which is the amount of bunch compression needed for a DC beam, this gives the drift distance needed from buncher to first rf gap.

For the case mentioned above, the TR-30, assuming a phase acceptance of $72^\circ$, $r_0=3$\,mm. At 1\,mA, $L_0=74$\,mm and at 2.5\,mA, it is 47\,mm. Then the buncher distance is resp.\ 37\,cm and 23\,cm and requires it to be within the heavily congested location inside the magnet pole yoke where there also is the optics dedicated to focusing the beam into the inflector.

Scaling up to higher injection energy alleviates this in a number of ways. Note that $V_0\propto\beta^2$ and $r_0\propto\beta\lambda_{\rm rf}$. Thus $L_0\propto V_0^{5/4}\lambda_{\rm rf}$. The TRIUMF cyclotron with $V_0$ larger by a factor 12, and $\lambda_{\rm rf}$ by a factor 3 gives a factor 60 length scale increase, and this is consistent with the current buncher placement at a convenient location $20$\,m from injection. Further, for higher $V_0$, the inflector and the central region can be larger, reducing the loss density, and lastly the erosion rate is intrinsically lower at higher energy, as discussed next.

For the other hydrogen molecules H$_2^+$, H$_3^+$ ($A=2,3$) as considered here, there isn't a simple way to compare the space charge effect on bunching. For fixed injection energy per charge $V_0$, magnetic field and harmonic number, $L_0$ increases by a factor of only $A^{1/4}$. But then the rf frequency decreases by a factor $A$, and the first turn orbit radius increases by a factor $A^{1/2}$. 

Maintaining the orbit geometry would require both $V_0$ and the central field $B_c$ to be increased by a factor $A$ as well. In that case, $L_0$ is larger by factor $A^{1/2}$. However, in designs where either $V_0$ or $B_c$ is at a design limit, this is an unfair comparison. It is understood of course that for given $\Ia$, proton rate increases as $A$.

\FloatBarrier
%\section{Losses on Residual Gas}
%The point I want to make: it is a limiting factor in cyclotron with internal ion sources, because of the insufficient differential pumping. In a cyclotron with external ion source, it is usually not an issue. This was verified experimentally in the case of TRIUMF 500\,MeV cyclotron by creating a controlled leak using different gases (BL1A). We recorded the corresponding total amount of beam loss, on dedicated spill monitors: the data is presented in~\cref{fig:spill}. By extrapolating these curves we can estimate the amount of beam that would be lost if we had a perfect vacuum in the tank. Since we normally run with a vacuum around $2\times10^{-8}$\,Torr, you can see on this figure that the contribution of the imperfect vacuum to beam loss is negligible.
%
%\begin{figure}
%    \begin{center}
%        \includegraphics[width=0.75\textwidth]{figure/Spills_vs_tkvacuum_H2HeN2.pdf}
%        \caption{Measured beam loss in TRIUMF 500\,MeV cyclotron, while varying the tank vacuum by created controlled leaks. At the time of the measurement, the machine was set up to send simultaneously $\sim$90\,$\upmu$A at 100\,MeV to beamlines 2C and $\sim$75\,$\upmu$A at 480\,MeV to beamline 1A + 2A.  The measurement was repeated 3 times with 3 different gases for the controlled leak: nitrogen, hydrogen and helium.}\label{fig:spill}
%    \end{center}
%\end{figure}
%
%\FloatBarrier
\section{Central Region Erosion}
When a target is bombarded by an ion beam, the incoming ions tend to ``kick" some surface atoms out of the target. This process is called sputtering. Sputtering is highly undesired in most accelerators as it causes permanent damage to the electrodes and frequent replacement of parts becomes necessary. As well, the sputtered atoms can build a conductive coating on insulators. This is especially true at the central region of a compact cyclotron, where most often a large fraction of the injected beam lies outside the phase acceptance and is dumped on metal surfaces. As well, space is confined and vertical focusing poor as discussed in section \ref{sec:phase_acceptance}.

\begin{figure}
    \begin{center}
        \includegraphics[width=0.7\textwidth]{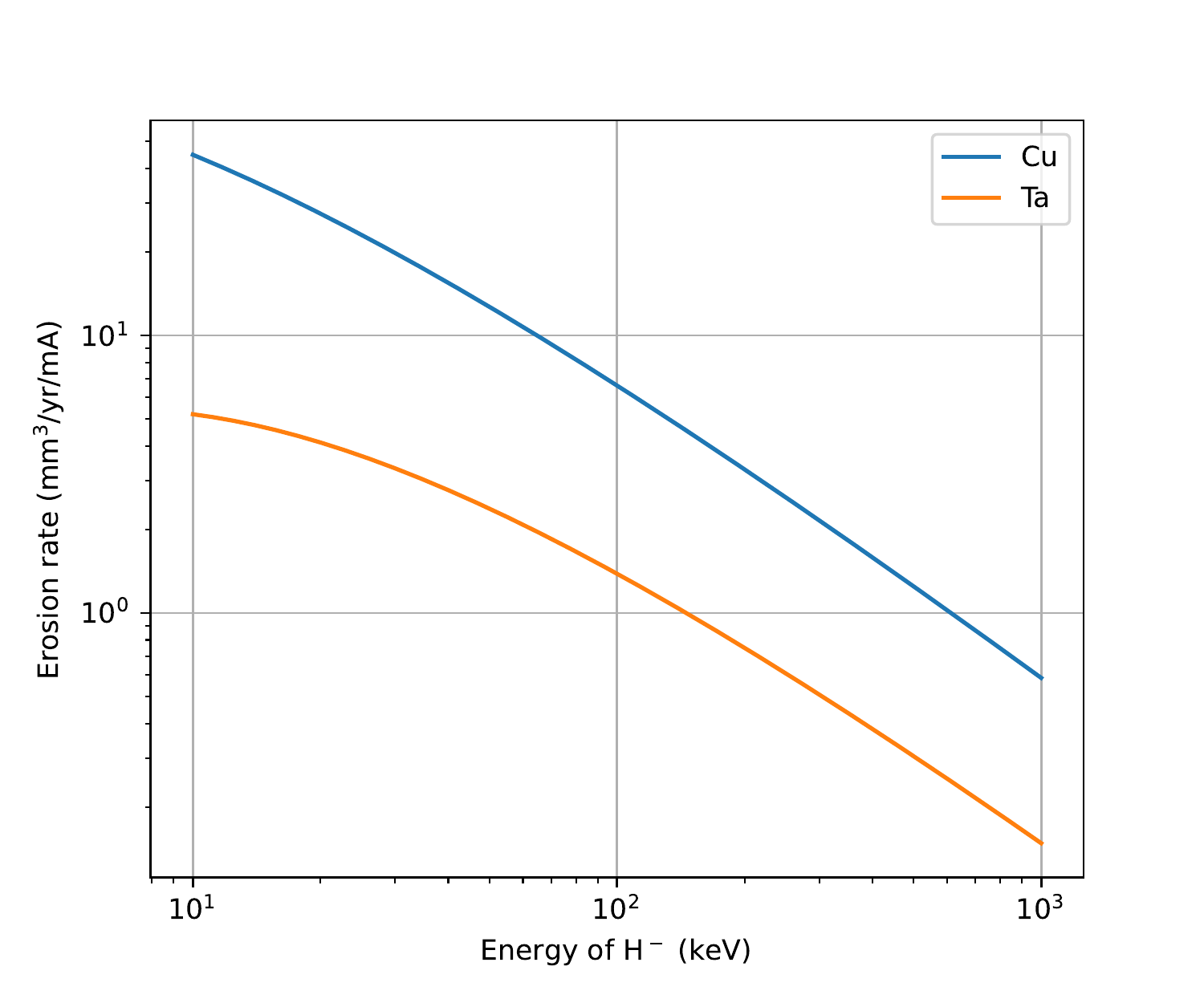}
        \caption{Sputtering erosion rate per year per mA of protons plotted for two different materials: copper and tantalum.}\label{fig:erosion_rate}
    \end{center}
\end{figure}

The surface sputtering rate is dependent on the type of projectile ions and the target elements. It can be estimated using the empirical formula given by Y. Yamamura et. al. \cite{yamamura1996sputter}:
\begin{equation}
Y(E)=0.042 \dfrac{Q \alpha}{U_s} \dfrac{S_n(E)}{1+\Gamma k_e \epsilon^{0.3}} \left( 1-\sqrt{\dfrac{E_{\mathrm{th}}}{E}} \right)^s
\label{eq:sputter1} 
\end{equation}
where $Q=1$, $\Gamma=0.728$ and $s=2.5$ are dimensionless fit parameters for Cu, $U_s=3.49 $ eV is the surface binding energy of Cu; $\alpha=4.29$ is the mass ratio parameter; $k_e=4.71$ is the Lindhard electronic stopping coefficient of H$^-$ ions in Cu target; $E_{\mathrm{th}}=61.9$ eV is the sputtering threshold; $\epsilon=0.000342 E$ (eV) is the reduced energy; $S_n=0.676\ s_n(\epsilon)$ is the nuclear stopping cross section in units of eV \AA$^2$/atom, $s_n=\dfrac{3.441\sqrt{\epsilon} \ln{(\epsilon+2.718)}}{1+6.355\sqrt{\epsilon} + \epsilon(6.882\sqrt{\epsilon} - 1.708)}$ is the reduced nuclear stopping power based on the Thomas-Fermi potential \cite{yamamura1996sputter}. Assuming the cyclotron is operating continuously for a year, the sputtering loss rate of Cu is shown in Fig.\,\ref{fig:erosion_rate}.

The TRIUMF cyclotron has injection energy of 300\,keV. Reading from the graph, the erosion rate is $\sim2$\,mm$^3$/yr/mA. It runs continuously for 8 months/yr at about 200\,$\upmu$A but has a very efficient bunching system that loses only about $30\%$ of the beam that lies outside the phase acceptance and is dumped on the centre post. This amounts to less than 1\,mm$^3$ in a year, and indeed the beam absorber that intercepts the lost beam has not required service in 20 years.

\begin{figure}
    \begin{center}
        \includegraphics[width=0.7\textwidth]{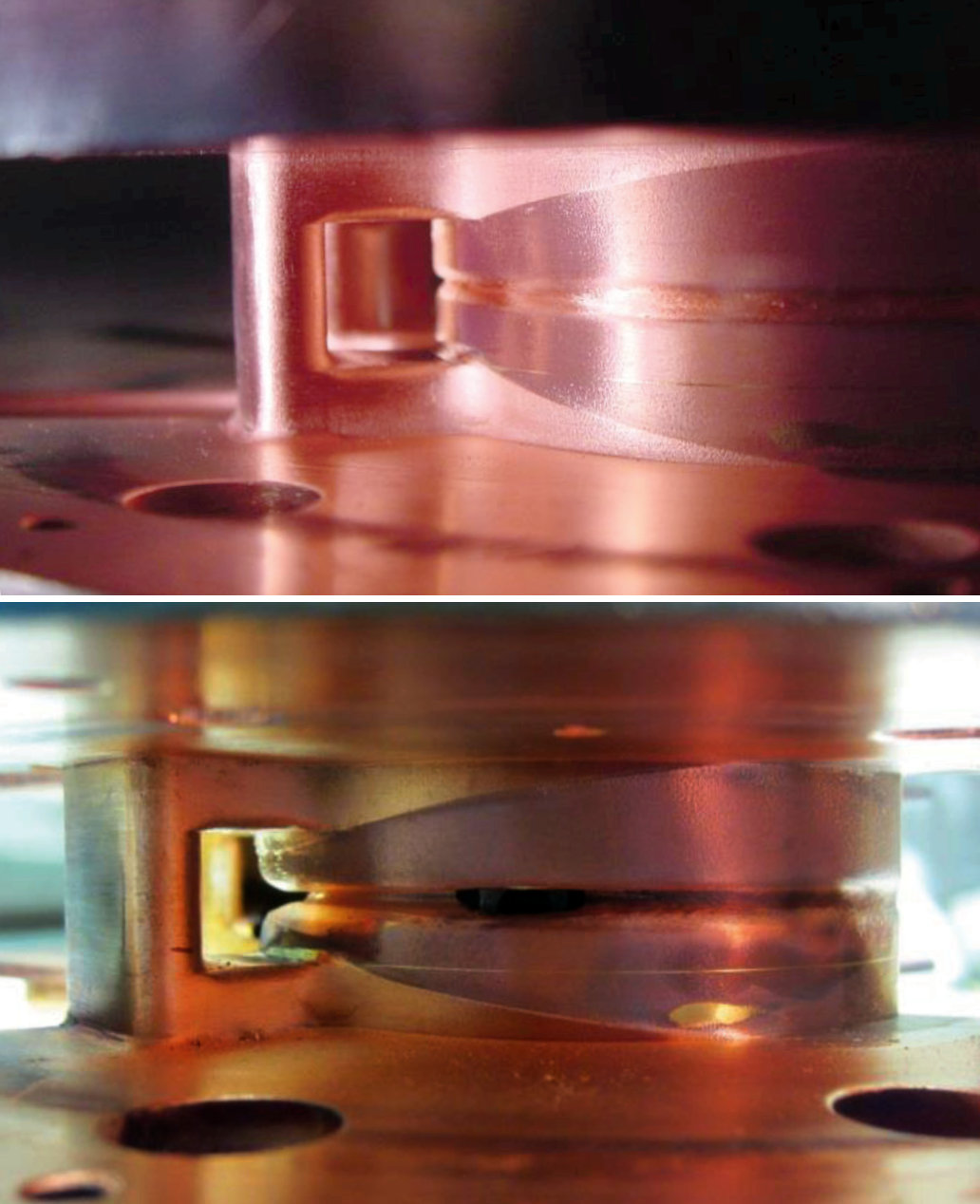}
        \caption{Erosion of the TR-30 centre region. Both photos are of the copper centre post (under slightly different lighting conditions), and the erosion is due to particles lost on the first turn. The top photo is after two years of continuous running, while the bottom is after ten years. Note the slight groove in the top photo that is due to sputtering from the particles that have not gained enough energy to clear the post. After ten years, this groove has become deep enough to cut completely through the copper block. For scale, the aperture height is 1\,cm.}\label{fig:erosion3}
    \end{center}
\end{figure}

Smaller cyclotrons such as the TR-30 have a more significant central region erosion than the larger TRIUMF cyclotron due to 3 reasons: (1) A lower injection energy (by 12 times) giving an order of magnitude higher erosion rate, (2) higher current loss of about factor 100 because of the impossibility of bunching efficiently, (3) a much smaller central region where the erosion occurs. An example of the erosion at the central region of TR-30 is shown in Fig.\,\ref{fig:erosion3}. In specific numbers, that cyclotron operates for $\sim 8$ months of the year, with a source output of $\sim 8$\,mA of which only $\sim1$\,mA survives the first turn. At 25\,keV, the rate is 15\,mm$^3$/(mA.yr), so this predicts 700\,mm$^3$ over the 10-year time span.  This is compatible with the photograph. (See \cite{cojocaru2022long} for further details.)

As any physical deterioration of the metal surfaces tends to reshape the beam's surroundings and the performance of the rf system, it is important to consider the sputtering loss when deciding the operational beam intensity. In addition, the type of electrode used also affects the sputtering erosion rate. For instance, Fig.\,\ref{fig:erosion_rate} shows that the erosion rate of Ta is almost an order of magnitude less than that of Cu. Therefore, using an electrode made from a high $Z$ material, such as Tantalum, could potentially alleviate the erosion, and further allowing a higher operational beam intensity and/or longer times between major maintenance.

Regarding other hydrogen molecules, note that the erosion rate is proportional to the rate of proton delivery, not charge delivery. Thus at for example, one electrical milliAmpere of H$_2^+$, while delivering twice the protons from the cyclotron, will also erode the centre region components twice as quickly as 1\,mA H$^-$. Worse, the protons impinging upon the copper surfaces do so at half the energy for a given extraction voltage $V_0$, since the kinetic energy is shared equally between the protons, and according to the graph, the erosion rate is roughly proportional to the inverse of the kinetic energy. Thus, for a given proton delivery rate, the heavier hydrogen molecules would erode centre region components more quickly than the H$^-$ delivery system would.

%\begin{figure}
%    \begin{center}
%        \includegraphics[width=0.75\textwidth]{figure/IMG_1804s}
%        \caption{Picture shows erosion of the TR-30 inflector. Judge scale by the technician's gloved fingers. This devices bends the beam from the vertical injection line onto the median plane. Note the eroded groove.}\label{fig:erosion1}
%    \end{center}
%\end{figure}
 
\FloatBarrier
\section{Conclusion}
Compact H$^-$ cyclotrons have been workhorses for isotope production for some years now. They have practical limits on the order of 1\,mA. Higher currents are achievable but at the cost of either more frequent maintenance, or an increase in the injection energy. Lorentz stripping limits such cyclotrons to around 70\,MeV~\cite{koay2022lorentz}. Higher energies with the same advantages of the very large phase acceptance 
%needed to retain mA-level intensities, 
will still only be possible with stripping extraction. This has caused a search for other strippable ion species. H$_2^+$ and H$_3^+$ are candidates, but because of the way the sputtering rate scales with the particle velocity, they appear only viable if injection energy is significantly increased.

\section{Acknowledgements}
We thank the anonymous reviewer who helped improve the work and proposed an exact expression for~\cref{eq:ttilde}, which we had previously only been able to express in the form of a truncated Taylor series.

\FloatBarrier
% \bibliographystyle{ieeetr}
% \bibliographystyle{unsrt}
% \bibliography{bib}

\end{document}